\documentstyle[sprocl]{article}

\def\be{\begin{equation}}
\def\ee{\end{equation}}
\def\bea{\begin{eqnarray}}
\def\eea{\end{eqnarray}}

\begin{document}

\title{Continuous space-time symmetries in a lattice field theory
\footnote{To appear in
\underline{Statistical Physics on the Eve of The Twenty-First Century},
a fetschrift for James B. McGuire.}}

\author{ H.B. THACKER }

\address{Dept. of Physics,\\ University of Virginia,\\ Charlottesville, VA
22901} 


\maketitle\abstracts{For purposes of regularization as well as
numerical simulation, the discretization of Lorentz invariant
continuum field theories on a space-time lattice is often convenient.
In general this discretization destroys the rotational or 
Lorentz-frame independence
of the theory, which is only recovered in the continuum limit.
The Baxter 8-vertex model may be regarded as a particular discretization of
a self-interacting massive Dirac fermion field theory in two 
dimensions (the massive
Thirring model). Here it is shown that, in the 8-vertex/massive Thirring
model, the Lorentz-frame independence of the continuum theory remains
undisturbed on the lattice. The only effect of the discretization
is to compactify the manifold of Lorentz frames. The relationship
between this lattice Lorentz symmetry and the Yang-Baxter relations 
is discussed.
} 
\section{Introduction}
\label{Introduction}

The early work of Jim McGuire on the delta-function \cite{McGuire} 
gas was the
first analysis to focus on the essential feature of consistent 
factorization of
3-body scattering amplitudes into products of 2-body amplitudes. 
Subsequent
developments of integrable systems technology have generalized and 
enshrined this 
idea in the form of a generic algebraic structure, the Yang-Baxter
relations \cite{Yang,Baxter8v}, which is common to a 
wide variety of solvable models.
The recognition that the transfer matrix methods used by Baxter
to solve the 8-vertex model arose from precisely the same 
algebraic
structure (this time as a statement about vertex Boltzmann weights
instead of scattering amplitudes) led to the quantum inverse 
scattering
method, an elegant unification of Yang-Baxter relations, the
Bethe ansatz, and classical
soliton methods.\cite{Faddeev,HBTRMP} In spite of all these developments, the physical
nature of the symmetries imposed by the Yang-Baxter relations is
still somewhat obscure. In this paper, I would like to discuss a 
viewpoint
on integrable models and Yang-Baxter relations 
which arises from a lattice construct
called a corner transfer matrix (CTM), invented and developed by
Rodney Baxter in the late 1970's \cite{BaxterCTM}. I will briefly review
some arguments which expose a connection between the CTM and Yang-
Baxter
relations of the 8-vertex model and the question of how space-time 
symmetries (or their remnants) are realized in a lattice field 
theory.\cite{HBTCTM,Itoyama,Frahm} 
Simply stated, the corner transfer matrix is the lattice 
generalization
of a Lorentz boost (or Euclidean rotation) operator. The 
miraculous
properties of the CTM (which follow from the Yang-Baxter 
relations)
are a manifestation of the fact that the
lattice theory supports an exact, continuous analog of the Lorentz
symmetry of the continuum theory. The obvious question of how a
continuous space-time symmetry can survive in any sense on the 
lattice
is answered by the introduction of the elliptic function or
``lattice rapidity'' parametrization of momentum space. 

To understand the nature of lattice rapidity, 
let us first recall the role of rapidity in a Lorentz invariant 
continuum theory in one spatial dimension.
The energy-momentum dispersion relation
$\omega(p) = \sqrt{p^2+m^2}$ for a relativistic particle  
of mass $m$, may be
parametrized by introducing a rapidity variable $\alpha$,
\begin{equation}
p = m\sinh\alpha,\;\;\;\; \omega = m\cosh\alpha
\end{equation}
The rapidity variable ``uniformizes'' momentum space, in the
sense that the relativistic phase space volume element becomes
\begin{equation}
\label{eq:rapidity}
\frac{dp}{2\omega(p)} = \frac{1}{2}d\alpha
\end{equation}
Rapidity also uniformizes momentum space from a dynamical point of 
view.
Specifically, Lorentz invariance implies that invariant
scattering amplitudes depend only on relative rapidity variables,
The scattering amplitudes are unchanged by a uniform shift of all 
the rapidity
variables in the scattering state, which is equivalent to a change
of the observer's Lorentz frame. As I discuss here, the 8-vertex
model and $XYZ$ spin chain have the amazing property that all of 
this
structure carries over undisturbed to the the lattice
theory. The only change introduced by the lattice is that momentum
space is compactified in the real rapidity direction. The 
uniformity of
rapidity space associated with Lorentz invariance remains 
unchanged.
For example, two-body phase shifts and Bethe ansatz kernels 
depend only on the relative rapidity of the colliding spin waves 
(leading to the ``difference kernel'' form of the Bethe ansatz
equations \cite{Baxter8v}).
Note that the periodicity
of the continuum rapidity parametrization (\ref{eq:rapidity}) in the 
imaginary direction $\alpha\rightarrow \alpha+2\pi i$ corresponds
to periodicity under Euclidean
rotations by $2\pi$.. In the lattice theory, the rapidity 
parametrization is given in terms of
doubly periodic elliptic functions, combining the periodicity 
under Euclidean
rotations (imaginary rapidity) 
with the lattice periodicity of  boosts (real rapidity) corresponding to
momentum shifts by $2\pi/a$ where
$a=$ lattice spacing. To see this structure in a simple context,  
in Section 3 I
will look specifically at the single particle eigenmode operators 
of the XY spin chain,
\begin{equation}
H = -\frac{1}{2}\sum_j\left[\sigma^x_j\sigma^x_{j+1}+k\sigma^y_j\sigma^y_{j+1}
\right]
\end{equation}
which corresponds to the free fermion case of the Thirring model.
The eigenmodes of $H$ can be
parametrized in terms of lattice rapidity variables labelling the
momenta of the spin waves. These variables are exactly 
analogous to the continuum rapidity in a relativistic theory. 
The corner transfer matrix, when applied to a spin-chain 
eigenstate, 
produces another eigenstate with a shifted rapidity variable. This 
is
completely analogous to the Lorentz boost operator in a continuum 
theory,  
which sweeps through the set of states of different total momentum 
corresponding
to the same physical state in different Lorentz frames. In the 
continuum case,
the states can be classified in terms of irreducible representations of the 
2-dimensional
Poincare algebra,
\begin{equation}
\label{eq:Poincare}
[\tilde{K}, \tilde{H}] =  \tilde{P},\;\;\;  
[\tilde{K},\tilde{P}]=\tilde{H},\;\;\; [\tilde{H},\tilde{P}]=0
\end{equation}
where $\tilde{H}, \tilde{P},$ and $\tilde{K}$ are the Hamiltonian, the total momentum
operator,
and the boost generator, respectively.
In the 8-vertex model, the Poincare algebra is replaced by the 
lattice boost
generator $K$ (which is given by the log of the CTM, see below) 
and an infinite tower of 
commuting 
conserved quantities $H_n$ 
\begin{equation}
[H_n,H_m] = 0
\end{equation}
which generalize the role of $\tilde{H}$ and $\tilde{P}$. Instead of closing as in
the continuum 
algebra, repeated commutation by $K$ walks up the tower of 
conserved
quantities, \cite{HBTCTM}
\begin{equation}
[K, H_n] = H_{n+1}
\end{equation}
The lowest member of this heirarchy $H_1\equiv H$ is the  
nearest-neighbor $XYZ$ spin chain Hamiltonian, with the higher
conserved operators $H_n$ involving interactions ranging up to
n$th$ nearest neighbor.

It was shown long ago by Alan Luther\cite{Luther} that the continuum theory 
obtained 
by approaching the critical point of the
8-vertex model is equivalent to the massive Thirring model,
a self-interacting, relativistic Dirac fermion theory. 
Perforce we may view the 8-vertex model as the result
of taking the continuum massive Thirring model and ``putting it on
a lattice,'' i.e. in some way discretizing the fermionic variables
of that theory. The direct connection between the vertex Boltzmann 
weights
and the two-dimensional action of the Dirac
fermion theory would be of great interest, but this connection has 
not yet 
been fully clarified. The relationship between 
the two models 
is most easily discussed in the transfer matrix
or Hamiltonian framework. It can be shown that the 
spin-chain Hamiltonian may be directly transformed into a lattice 
Dirac Hamiltonian\cite{Lat98} as discussed in Section 4. 
The comparison of the spin-chain eigenmodes 
with those 
of the continuum Dirac Hamiltonian clearly exhibits the structure
imposed by the lattice Lorentz invariance embodied in the CTM 
formalism.

The direct transcription of the lattice spin Hamiltonian to a 
lattice
Dirac Hamiltonian allows us to address the interesting question of 
exactly how the spin chain constitutes a discretization of the 
Dirac field. 
The spin-chain fermion operators $c^x_j, c^y_j$ 
on site $j$ are obtained via a Jordan-Wiger
transformation of the local Pauli spin matrices,
\begin{equation}
\label{eq:JW}
c^x_j = \sigma^x_j\prod_{l<j} \sigma^z_l,\;\;\;
c^y_j = \sigma^y_j\prod_{l<j} \sigma^z_l
\end{equation}
This gives us two real fermion operators on each site. A lattice
Dirac field consists of a two-component complex Dirac spinor, i.e. 
four
real fermion operators on each site. In a particular 
representation of Dirac matrices
(with $\gamma^1$ diagonal), the $x$ and $y$ labels of the spin-
chain
fermions $c^x_j$ and $c^y_j$ correspond to upper and lower 
components
of the Dirac spinor. However, the real and imaginary parts of a 
single Dirac spinor
component correspond to nearest neigbor combinations of spin-chain 
operators. (See Section 4.)
As a result, the vector charge symmetry of the Dirac field, 
corresponding
to local phase rotations of the complex spinor components, is 
expressed
in terms of spin-chain operators by a non-local mixing of nearest
neighbor pairs. The associated conserved charge is thus not 
locally defined
on the spin lattice. (In fact, the conserved vector charge on the 
lattice is
the kink number associated with the Bethe ansatz for this model, 
and
introduced via Baxter's SOS transformation
\cite{Baxter8v2,Jones}). On the other hand, the chiral symmetry of 
the theory in
the limit of zero fermion mass is related to a very simple and 
obvious
symmetry of the spin chain. The massless fermion theory 
corresponds
to a spin-spin interaction which is isotropic in 
the $\sigma^x-\sigma^y$ plane, 
i.e.
has equal coefficients for the $\sigma^x_j\sigma^x_{j+1}$
and $\sigma^y_j\sigma^y_{j+1}$ terms. The global chiral symmetry 
which arises
in the massless fermion theory corresponds to the symmetry of
the isotropic spin chain under global rotations in the 
$\sigma^x-\sigma^y$ plane. Some 
time ago,
I argued that the chiral and Lorentz properties of the Heisenberg 
spin
chain and 8-vertex model made this a particularly interesting 
model 
for studying properties of chiral lattice fermions.\cite{Lat95} 
It can be shown \cite{Lat98} that the free lattice Dirac
Hamiltonian obtained from the $XY$ 
spin-chain is a Wilson-Dirac Hamiltonian operator with Wilson 
parameter
$r=1$ and hopping parameter $\frac{1}{2}k$, and that,
for the massless case $k=1$, it 
satisfies a one-dimensional version
of the Ginsparg-Wilson relations\cite{Ginsparg}:
\begin{equation}
\{\gamma^5,{\cal D}\} = {\cal D}\gamma^5{\cal D}
\end{equation}
where ${\cal D}=\gamma^0{\cal H}$, and ${\cal H}$ is the lattice
Dirac Hamiltonian.
Furthermore, it can be shown \cite{Horvath} that this particular
Wilson-Dirac operator is unique in that it is the only lattice
Dirac operator in one-dimension involving a finite number of
nearest-neighbor hopping terms which satisfies the Ginsparg-
Wilson relations.
This may suggest a connection between
vertex models and the form of lattice chiral symmetry that is 
embodied
in the Ginsparg-Wilson  relations.  
This problem is currently under investigation. \cite{Lat98}

\section{Lattice anisotropy, commuting transfer matrices, and
Lorentz frame independence}

To introduce the discussion of lattice Lorentz invariance, let me 
describe in generic
terms how this Lorentz invariance is related to the Yang-Baxter 
relations.
In the row-to-row transfer matrix (or quantum inverse) formalism, a 
vertex is
represented by a $2\times 2$ matrix of spin operators 
$V_j(\alpha)$ where each element of this matrix contains Pauli 
matrices
acting on spin $j$. The vertex is a function of
a rapidity-like anisotropy parameter $\alpha$ (sometimes called the
spectral parameter because of its role in the quantum inverse method), which determines the 
Boltzmann weights
via Baxter's elliptic function parametrization. (The other two 
parameters in the 
vertex weights are essentially the mass and coupling of the Thirring 
model and are 
treated as fixed constants.) The transfer matrix $T(\alpha)$ is 
then
given by a row of vertices of the form
\begin{equation}
\label{eq:TM}
T(\alpha) = Tr\left(\prod_{j=-L}^{j=L} V_j(\alpha)\right)
\end{equation}
where the trace and product are over the $2\times 2$ matrix 
space (horizontal arrows of the vertex model). 
Although I have implicitly assumed spatial periodic boundary
conditions by taking the trace in (\ref{eq:TM}),
in what follows, I will effectively assume that the chain of spins 
stretches from
$-\infty$ to $\infty$ and ignore the subtleties associated with
boundary terms. (In the CTM framework, issues associated with
the limit of infinite spatial volume are replaced by issues of
analytic continuation to complex momentum or rapidity (c.f (\cite{Itoyama}).
The Yang-Baxter relations for this model are trilinear algebraic
relations among vertices $V_j(\alpha)$.
It is easy to show that the Yang-Baxter relations, in the limit 
where two of the
rapidty parameters involved are nearly equal, reduce to a simple 
statement
giving the commutator of the nearest-neighbor spin-chain 
Hamiltonian
term ${\cal H}_{j,j+1}$ and the product of the two vertices at 
sites $j$
and $j+1$, namely
\begin{equation}
\label{eq:YBE}
[{\cal H}_{j,j+1}, V_j(\alpha)V_{j+1}(\alpha)] = 
V_j(\alpha)\left(\frac{\partial}{\partial\alpha}V_{j+1}(\alpha)\right) 
- \left(\frac{\partial}{\partial\alpha}V_j(\alpha)\right) 
V_{j+1}(\alpha)
\end{equation}
In the following discussion, the properties of two important 
lattice operators 
constructed from ${\cal H}_{j,j+1}$
are relevant. One is the standard Heisenberg spin chain 
Hamiltonian,
\begin{equation}
H = \sum_j {\cal H}_{j,j+1}
\end{equation}
Most of the following discussion
applies to the general symmetric 8-vertex model, equivalently, the 
fully anisotropic
XYZ spin-chain, for which
\begin{equation}
{\cal H}_{j,j+1} = -\frac{1}{2}\left[\sigma^x_j\sigma^x_{j+1} + 
k\sigma^y_j\sigma^y_{j+1} + \Delta \sigma^z_j\sigma^z_{j+1}\right]
\end{equation}
In Section 3, I will analyze the free fermion case $\Delta=0$ 
more completely.
The other operator essential to the discussion is the lattice 
boost generator,
which is given by the first moment of the same Hamiltonian 
density,
\begin{equation}
\label{eq:boost2}
K = \sum_j j{\cal H}_{j,j+1}
\end{equation}
(Note that this is completely analogous to the continuum boost 
generator which
is given, at $t=0$, by the first moment of the Hamiltonian 
density, 
$K = \int x{\cal H}(x) dx$.)  
An important property of $K$ is that, because the $j=0$ term in the
sum (\ref{eq:boost2}) vanishes, it separates into two commuting 
operators which act separately on the left and right half-chain,
\begin{equation}
K = K^> +K^<
\end{equation}
As Baxter showed in his original work, the corner transfer
matrix in the infinite volume limit is, up to an overall 
constant, exactly the exponential of the operator $K^>$
or $K^<$ (sweeping out the left- and right-hand corners of
the lattice, respectively).

From the Yang-Baxter commutator
(\ref{eq:YBE}) it follows that $K$ has the following commutator 
with
the transfer matrix:
\begin{equation}
[K,T(\alpha)] = \frac{\partial}{\partial\alpha} T(\alpha)
\end{equation}
Thus $K$ generates a shift of the rapidity (anisotropy) 
parameter,
\begin{equation}
e^{i\beta K}T(\alpha)e^{-i\beta K} = T(\alpha + \beta)
\end{equation}
In a sense, we can interpret $T(\alpha)$ for a particular value of 
$\alpha$ as the transfer matrix in a particular Lorentz frame. The
boost operator $K$ generates Lorentz transformations from one 
frame
to another by shifting the rapidity of the Lorentz frame. Among 
other things,
this explains why transfer matrices with different values of $\alpha$ 
commute with each other (the famous result that led Baxter to
his solution of the model),
\begin{equation}
[T(\alpha),T(\alpha')] = 0
\end{equation}
This follows from the fact that any two observers of the same 
theory in two different frames will construct the
same set of  Hamiltonian or transfer matrix eigenstates. Thus, 
$T(\alpha)$
and $T(\alpha')$ are simultaneously diagonalizable.
In the continuum, this follows from the fact that $H$
and $P$ commute (because a boost simply mixes $H$ and $P$). 
On the lattice, it requires an infinite
number of commuting conserved quantities.
If the rapidity parameter which appears in the vertex weights is 
taken to be imaginary,
it is most appropriately interpreted as a lattice
anisotropy parameter which determines the relative scale between 
the space
and time directions. The Heisenberg spin-chain Hamiltonian is the 
first nontrivial
term in an expansion of $\log\;T(\alpha)$ in powers of $\alpha$.
Here $\alpha\rightarrow 0$ corresponds to the limit in which
the lattice spacing in the time direction goes to zero.
In this sense, the Heisenberg spin-chain is the time-continuum 
Hamiltonian formulation
of the two-dimensional 8-vertex model.. 

\section{Lorentz transformation of lattice spin waves and Dirac
fermions}

To exhibit some of the simpler consequences of lattice Lorentz 
invariance, I will look at
the eigenmode operators for the $XY$ Hamiltonian. 
For comparison, we first review the Lorentz transformation
properties of eigenmodes of the free massive continuum Dirac Hamiltonian.
In a Hamiltonian framework, the most direct statement of Lorentz invariance
is in terms of the action of the boost operator on the eigenstates of the
Hamiltonian. It is this formulation that generalizes most directly to the
lattice theory. We will work in Minkowski space, 
taking $\left(\gamma^0\right)^2=1$, $\left(\gamma^1\right)^2=-1$, and
$\gamma^5=\gamma^0\gamma^1$. The continuum Dirac Hamiltonian is
\begin{equation}
\tilde{H} = \frac{1}{2}\int dx\;\left[\bar{\psi}\left(-i\gamma^1\partial_1 + m\right)\psi
+ {\rm h.c.}\right]
\end{equation}
The comparison with the $XY$ spin chain is facilitated by
choosing a particular representation of gamma matrices for
which $\gamma^1$ is diagonal. Specifically, I will take
$\gamma^0=\sigma^x,$ $\gamma^1= i\sigma^z$, and 
$\gamma^5=\sigma^y$, where the $\sigma$'s are the standard
Pauli matrices. In this basis, we can derive
the Hamiltonian equations for the free Dirac field in
terms of the fourier transformed field,
\begin{equation}
\tilde{\psi}(p) = \int dx\;e^{-ipx}\psi(x)
\end{equation}
The spinor components satisfy
\begin{eqnarray}
\left[\tilde{\psi}_1(p),\tilde{H}\right] = (m-ip)\tilde{\psi}_2(p) \nonumber\\
\left[\tilde{\psi}_2(p),\tilde{H}\right] = (m+ip)\tilde{\psi}_1(p)
\end{eqnarray}
Defining
\begin{eqnarray}
\label{eq:bops}
b_1(p) = (m+ip)^{\frac{1}{2}}\tilde{\psi}_1(p) \nonumber \\
b_2(p) = (m-ip)^{\frac{1}{2}}\tilde{\psi}_2(p)
\end{eqnarray}
the Hamiltonian equations reduce to 
\begin{eqnarray}
\left[b_1(p),\tilde{H}\right] = \omega(p)b_2(p) \nonumber\\
\left[b_2(p),\tilde{H}\right] = \omega(p)b_1(p)
\end{eqnarray}
where
\begin{equation}
\omega(p) = \left(p^2+m^2\right)^{\frac{1}{2}}
\end{equation}
Thus, the eigenmodes of $\tilde{H}$ are
\begin{equation}
b^{\pm}(p) = b_1(p)\mp b_2(p)
\end{equation}
which satisfy
\begin{equation}
\left[\tilde{H},b^{\pm}(p)\right] = \pm \omega(p)b^{\pm}(p)
\end{equation}
Since $b^+(p)$ and $b^-(p)$ carry the same fermionic
charge but have opposite energy, they must be interpreted
as particle creation and antiparticle annihilation operators,
respectively.

Next, consider the action of the continuum Lorentz boost 
operator
\begin{equation}
\label{eq:boost}
\tilde{K} = \frac{1}{2}\int dx\;x\;\left[\bar{\psi}\left(-i\gamma^1\partial_1 
+ m\right)\psi + {\rm h.c.}\right]
\end{equation}
The commutators of $\tilde{K}$ with the Dirac field may be computed in a
similar way to the Hamiltonian, with the additional factor of $x$
in the integrand of (\ref{eq:boost}) giving rise to a derivative
with respect to momentum. The commutators with $\tilde{K}$ are
particularly simple when expressed in terms of the operators defined
in (\ref{eq:bops}),
\begin{eqnarray}
\left[b_1(p),\tilde{K}\right] = i\omega(p)\frac{\partial}{\partial p}b_2(p)\\
\left[b_2(p),\tilde{K}\right] = i\omega(p)\frac{\partial}{\partial p}b_1(p)
\end{eqnarray}
Introducing the continuum rapidity, where $p=m\;\sinh\;\alpha$,
we see that the boost operator induces a uniform shift of the
rapidity of Hamiltonian eigenmodes,
\begin{equation}
\left[b^{\pm},\tilde{K}\right] = i\frac{\partial}{\partial \alpha}
b^{\pm}
\end{equation}
For comparison with the corresponding lattice results, it is
useful to note that the square root factors appearing in the 
expressions for the eigenmodes are entire functions of rapidity
\begin{equation}
\label{eq:Bogol}
(m\pm ip)^{\frac{1}{2}} = \sqrt{2m}\;\cosh\left(\frac{\alpha}{2}
\pm i\frac{\pi}{4}\right)
\end{equation}
Thus, we may write the eigenmodes entirely in terms of rapidity
\begin{equation}
\label{eq:contmode}
b^{\pm} = \sqrt{2m}\left[\cosh\left(\frac{\alpha}{2}+i\frac{\pi}{4}\right)
\tilde{\psi}_1(p(\alpha)) \mp \sinh\left(\frac{\alpha}{2}+i
\frac{\pi}{4}\right)\tilde{\psi}_2(p(\alpha))\right]
\end{equation}

Written in terms of
the fermionized spin operators (\ref{eq:JW}), the Hamiltonian 
becomes
\begin{equation}
H = -\frac{i}{2}\sum_j\left[c^x_{j+1}c^y_j + 
kc^x_jc^y_{j+1}\right]
\end{equation}
The procedure for diagonalizing this operator is well-known and 
consists
of a Fourier transform to momentum space followed by a Bogoliubov
transformation. Define the momentum-space fermion operators
\begin{equation}
a_{x,y}(z) = \sum_j z^j c^{x,y}_j
\end{equation}
It is easy to show that
\begin{eqnarray}
\left[H,a_x(z)\right]& = & \frac{i}{2}(z+kz^{-1})a_y(z) \\
\left[H,a_y(z)\right]& = &-\frac{i}{2}(z^{-1}+kz)a_x(z)\nonumber
\end{eqnarray}
Thus, if we define
\begin{eqnarray}
\label{eq:Bogol2}
B_x(z) & = (1+kz^2)^{\frac{1}{2}}\;a_x(z)\\
B_y(z) & = z(1+kz^{-2})^{\frac{1}{2}}\;a_y(z)
\end{eqnarray}
then
\begin{eqnarray}
\left[H,B_x(z)\right] = i\omega(z)B_y(z) \\
\left[H,B_y(z)\right] = -i\omega(z)B_x(z)
\end{eqnarray}
where we have defined the single-particle energy by
\begin{equation}
\label{eq:omega}
\omega(z) = \frac{1}{2}(1+kz^2)^{\frac{1}{2}}(1+kz^{-2})^{\frac{1}{2}}
\end{equation}
The Hamiltonian eigenmodes are given by
\begin{equation}
B^{\pm}(z) = B_x(z)\pm iB_y(z)
\end{equation}
which satisfy
\begin{equation}
\left[H,B^{\pm}(z)\right] = \pm\omega(z)B^{\pm}(z)
\end{equation}
Thus, $B^+(z)$ ($B^-(z)$) is a single particle creation 
(annihilation)
operator for a particle of energy $\omega(z)$. With a particular
choice of branch cuts for the square roots in (\ref{eq:omega}), 
the energy function
$\omega(z)$ is continuous and 
positive definite on the unit circle $|z| = 1$. The 
minima
of $\omega(z)$ are at $z=\pm i$, so we define the single-particle 
momentum $p$
to be given by $z=ie^{ip}$. (Here and elsewhere I take $a=1$, 
expressing
quantities like $p$ in lattice units.) In the continuum limit 
$p\rightarrow 0, k\rightarrow 1$,
$\omega(z)$ reduces to the continuum relativistic energy,
\begin{equation}
\omega(z)\rightarrow \sqrt{m^2+p^2}
\end{equation}
where the fermion mass is
\begin{equation}
ma = \frac{1}{2k}-\frac{1}{2}
\end{equation}
The commutators of the lattice boost operator $K$ with the spin chain
fermion operators $a_x(z)$ and $a_y(z)$ are easily computed in a
manner similar to the Hamiltonian commutators.
The action of the boost operator is very simply expressed in terms
of the Hamiltonian eigenmode operators. In terms of the spin chain
fermions, the boost generator is
\begin{equation}
K = -\frac{i}{2}\sum_j j\left[c^x_{j+1}c^y_j + kc^x_jc^y_{j+1}\right]
\end{equation}
By direct commutation one obtains
\begin{eqnarray}
\left[K,B_x(z)\right] = iz\omega(z)\frac{\partial}{\partial z}B_y(z)\\
\left[K,B_y(z)\right]= -iz\omega(z)\frac{\partial}{\partial z}B_x(z)
\end{eqnarray}
so that the eigenmode operators satisfy
\begin{equation}
[K,B{\pm}(z)] = z\omega(z)\frac{\partial}{\partial z}B^{\pm}(z)
\end{equation}
These commutators provide the essential statement of Lorentz
transformations on a lattice. Just as in the continuum, we may 
uniformize momentum space by defining a rapidity $\alpha$,
which is determined up to an overall constant by the differential 
relation
\begin{equation}
d\alpha = \frac{dp}{\omega(p)}=-i\frac{dz}{z\omega(z)}
\end{equation}
The solution to this is an elliptic function,
\begin{equation}
\label{eq:definealpha}
z(\alpha) = i\sqrt{k}\;{\rm sn}\frac{1}{2}(\alpha-\alpha_0)
\end{equation}
where ${\rm sn}$ is a Jacobian elliptic function of modulus $k$. 
The choice of
$\alpha_0$ will be made so that the lattice rapidity $\alpha$ 
corresponds to 
the ordinary rapidity in the continuum limit. This is accomplished 
by taking
\begin{equation}
\alpha_0 = 2\hat{K}+i\hat{K}'
\end{equation}
where $\hat{K}$ and $\hat{K}'$ are the complete elliptic integrals of modulus 
$k$ and $k'=\sqrt{1-k^2}$, respectively. ($\hat{K}$ and $\hat{K}'$ are the real and 
imaginary
elliptic quarter-periods.) 

Note that the Bogoliubov factors which appear in the Hamiltonian 
eigenmode 
operators (c.f. Eq. \ref{eq:Bogol2}) are also simply expressed in 
terms of 
elliptic functions:
\begin{eqnarray}
(1+kz^2)^{\frac{1}{2}}={\rm dn}\frac{1}{2}(\alpha-\alpha_0)\\
z(1+kz^{-2})^{\frac{1}{2}}=\sqrt{k} {\rm cn}\frac{1}{2}(\alpha-\alpha_0)
\end{eqnarray}
These two expressions are the analog of the continuum expressions
(\ref{eq:Bogol}). Thus, we can write the lattice eigenmodes as
functions of rapidity
\begin{equation}
B^+ = {\rm dn}\frac{1}{2}(\alpha-\alpha_0)\;\; a_x\left(z(\alpha)\right)
+i\sqrt{k}\;\;{\rm cn}\frac{1}{2}
(\alpha-\alpha_0)\; a_y\left(z(\alpha)\right)
\end{equation}
It is a simple exercise to show that the coefficients of $a_x$
and $a_y$ reduce to those of $\tilde{\psi}_1$ and $\tilde{\psi}_2$
in the continuum eigenmode (\ref{eq:contmode}). In taking the
continuum limit ($k\rightarrow 1$) of the elliptic functions, the
fact that the shift $\alpha_0$ goes to infinity in this limit must
be noted. A useful identity to use before taking the continuum limit
is
\begin{equation}
\sqrt{k}\;{\rm sn}(\beta-\hat{K}-i\frac{\hat{K}'}{2})
= \left(\frac{{\rm cn}\beta\;{\rm dn}\beta
+i(1-k){\rm sn}\beta}{{\rm cn}\beta\;{\rm dn}\beta
-i(1-k){\rm sn}\beta}\right)^{\frac{1}{2}}
\end{equation}
along with similar identities for ${\rm dn}$ and ${\rm cn}$.
To summarize, the lattice boost operator $K$ given by the first
moment of the spin chain Hamiltonian has exactly the same effect
on the lattice spin wave eigenmodes $B^+$ that the continuum
Dirac boost operator has on the eigenmodes of the Dirac Hamiltonian.
That is, it generates a shift of the rapidity variable.
This is the central manifestation of lattice Lorentz invariance
in the free fermion theory. I will not discuss the interacting
case $\Delta\neq 0$ in any detail, and some issues remain to
be investigated for this case. However, it is clear from corner
transfer matrix and Bethe ansatz results that the Lorentz invariance applies
to this case as well. For example, the two-body phase shift which
appears in the Bethe ansatz for the interacting case is Lorentz
invariant in the sense that it depends only on the relative rapidity
of the two colliding spin waves. For zero mass, the interacting
case reduces to the $XXZ$ spin chain. The boost properties of the
eigenmodes for this case have been investigated in detail.\cite{Frahm}.

\section{Spin-chain fermions and Dirac fermions}

The comparison of spin chain eigenmodes $B^{\pm}$ and Dirac 
eigenmodes $b^{\pm}$ in the
last Section would constitute a complete identification of the
two theories if the spin chain operators $a_x(z)$ and
$a_y(z)$ reduced to the Dirac fermion operators $\tilde{\psi}_1(p)$
and $\tilde{\psi}_2(p)$ in the continuum limit. However, 
the correspondence is not that simple, because of the fact that
$a_x(z)$ and $a_y(z)$ are the Fourier transforms of {\it real}
lattice fermions $c^x_j$ and $c^y_j$, and satisfy
\begin{equation}
\left(a_{x,y}(z)\right)^{\dag} = a_{x,y}(z^*)
\end{equation}
This contrasts with the Dirac field which is complex. The
states created by $b^+$ and $(b^-)^{\dag}$ carry opposite 
vector charge,
so that the particle spectrum consists of two distinct species
(particle and antiparticle). At first site, the 
reality of $a_x$ and $a_y$ would seem to indicate that
the spin chain contains only one species, and therefore
cannot constitute a Dirac fermion. The resolution of this
question, and the appearence of vector charge in the spin
chain, depends on the well-known doubling of the spectrum
associated with lattice fermions. The energy function 
$\omega(z)$ is an even function of $z=ie^{ip}$ and has two distinct
low-energy regions in the continuum limit, at $z=\pm i$.
To see how this doubled spectrum gets converted into the
charge of the fermion, let's look at the spin chain eigenmode
as a function of $z$,
\begin{equation}
B(z) \equiv B^+(z) = C(z)a_x(z) + izC(z^{-1})a_y(z)
\end{equation}
where
\begin{equation}
C(z) \equiv (1+kz^2)^{\frac{1}{2}}
\end{equation}
Because of the degeneracy under $z\rightarrow -z$, we may
"reduce the Brillouin zone" by defining eigenmodes which are
even functions of $z$ and allowing $p$ to go from $-\frac{\pi}{2}$
to $\frac{\pi}{2}$ instead of from $-\pi$ to $\pi$.
Define the eigenmodes in the reduced zone by
\begin{eqnarray}
B_1(z) = B(z) + B(-z)\\
B_2(z) = z^{-1}\left(B(z)-B(-z)\right)
\end{eqnarray}
Note that
\begin{eqnarray}
B_1(z) = C(z)a_x^e(z) +izC(z^{-1})a_y^o(z) \\
B_2(z) = z^{-1}C(z)a_x^o(z) + iC(z^{-1})a_y^e(z)
\end{eqnarray}
where $a_x^{e,o}$ are Fourier transforms over even and
odd sublattices,
\begin{equation}
a_{x,y}^{e,o}(z) = a_{x,y}(z) \pm a_{x,y}(-z)
\end{equation}
Since $B_1(z)$ and $B_2(z)$ in the reduced zone are
two independent creation operators with degenerate
energy, we may construct a symmetry transformation
which mixes $B_1$ and $B_2$. The corresponding
positively and negatively charged eigenmodes are
$B_1\pm iB_2$. From this construction, we can identify
the local lattice Dirac fields, which are given by
\begin{eqnarray}
\tilde{\Psi}_1(z) = \frac{1}{\sqrt{2}}\left(a_x^e(z)+iz^{-1}a_x^o\right)\\
\tilde{\Psi}_2(z) = \frac{1}{\sqrt{2}}\left(za_y^o(z)+ia_y^e(z)\right)
\end{eqnarray}
Equivalently, we may define the lattice Dirac fermion field
$\Psi^{1,2}_j$ at site $j$ (on a lattice with twice 
the lattice spacing of the original spin chain lattice) by
\begin{eqnarray}
\label{eq:latfermi}
\Psi^1_j = \frac{1}{\sqrt{2}}\left(c^x_{2j}+ic^x_{2j+1}\right)\\
\label{eq:latfermi2}
\Psi^2_j = \frac{1}{\sqrt{2}}\left(c^y_{2j}-ic^y_{2j-1}\right)
\end{eqnarray}
The vector fermion charge in terms of spin chain operators is thus,
\begin{equation}
\label{eq:charge}
Q = \sum_j\left[\Psi^{1 \dag}_j\Psi^1_j+\Psi^{2 \dag}_j\Psi^2_j\right]
= \sum_{j even}\left[c^x_{2j}c^x_{2j+1}+ c^y_{2j-1}c^y_{2j}\right]
\end{equation}
It is easy to show explicitly that $Q$ commutes with the $XY$
Hamiltonian,
\begin{equation}
[Q,H] = 0
\end{equation}
It is interesting to note that this conserved charge has appeared
in the literature on the 8-vertex model. In Baxter's original
papers on the Bethe ansatz for the 8-vertex model \cite{Baxter8v2},
he introduced an $SOS$ formulation of the model and an associated
conserved ``kink'' number. (The existence of this conserved number
of kinks is crucial for the formulation of the Bethe ansatz.)
This $SOS$ transformation was studied for the free fermion case
by Jones \cite{Jones}. There it was shown that the conserved kink number
associated with Baxter's $SOS$ transformation is precisely the 
operator $Q$ defined by (\ref{eq:charge}). 

Using the identification (\ref{eq:latfermi}-\ref{eq:latfermi2}),
the spin chain Hamiltonian may be directly transformed to a 
Wilson-Dirac Hamiltonian with Wilson parameter $r=1$ and 
hopping parameter $\frac{1}{2}k$,\cite{Lat98},
\begin{eqnarray}
H & = & \sum_j\left[\Psi^{1 \dag}_j\Psi^2_j+\Psi^{1 \dag}_j
\Psi^2_{j+1} + h.c \right] \\
& = & \sum_j\left[\bar{\Psi}_j\Psi_j +\frac{1}{2}k\left(
\bar{\Psi}_j\left(1+i\gamma^1\right)\Psi_{j+1}
+\bar{\Psi}_j\left(1-i\gamma^1\right)\Psi_{j-1}\right)\right]
\end{eqnarray}

\section{Conclusion}

The use of a space-time lattice to regularize relativistic quantum
field theories is now commonplace in both theoretical and numerical
investigations. For many nonperturbative questions, the lattice
formulation is the only well-defined cutoff scheme available.
Although there are typically an infinite number of different
lattice theories that correspond to a given continuum theory,
in practice one generally tries to retain in the lattice theory
as much of the symmetry of the continuum theory as possible.
The incorporation of exact global and vector-like gauge symmetries
on the lattice is usually not problematic, but fundamental
difficulties are encountered in the formulation of fermions
interacting with chiral gauge fields, as in the electroweak
Standard Model. Much effort has been devoted to understanding
the chiral lattice fermion problem, and there have been recent
promising developments.\cite{NN,Hasen,Neuberger,Luscher}. 
In a continuum field theory, the chiral structure of a fermion field
may be defined kinematically in terms of its Lorentz transformation
properties (i.e. the chiral components of the field are irreducible
under proper Lorentz transformations). In this paper, I have shown
that a two-dimensional Dirac fermion theory may be discretized as a
vertex model, and that this particular discretization retains
the full Lorentz symmetry of the continuum theory and merely compactifies
the manifold of Lorentz frames. Further investigation of the chiral
structure of this model may reveal some useful insights.

\section*{Acknowledgments}
I am grateful to I. Horvath for discussion of these and related issues.
This work was supported in part by the Department of Energy under
grant DE-FG02-97ER41027.

\end{document}